# Superconductivity onset above 60 K in ambient-pressure nickelate films

Guangdi Zhou[1,2†], Heng Wang[1,2†], Haoliang Huang[1,2†], Yaqi Chen[1], Fei Peng[1], Wei Lv[1], Zihao Nie[1], Wei Wang[1], Jin-Feng Jia[1,2,3], Qi-Kun Xue[1,2,4*], Zhuoyu Chen[1,2*]

[1]State Key Laboratory of Quantum Functional Materials, Department of Physics and Guangdong Basic Research Center of Excellence for Quantum Science, Southern University of Science and Technology, Shenzhen 518055, China
[2]Quantum Science Center of Guangdong-Hong Kong-Macao Greater Bay Area, Shenzhen 518045, China
[3]State Key Laboratory of Micronano Engineering Science, Tsung-Dao Lee Institute & School of Physics and Astronomy, Key Laboratory of Artificial Structures and Quantum Control (Ministry of Education), Shanghai Jiao Tong University, Shanghai 200240, China
[4]Department of Physics, Tsinghua University, Beijing 100084, China
[†]Equally contributed to this work.
[*]E-mail: chenzhuoyu@sustech.edu.cn; xueqk@sustech.edu.cn

## ABSTRACT

Ambient-pressure superconductivity in nickelates has been capped at an onset transition temperature ($T_c^{onset}$) of ~50 K, a value that remains lower than the cuprate (~133 K) and iron-based (~55 K) counterparts, despite the promise shown under high pressure. Here, we report ambient-pressure superconductivity onset at ~63 K in epitaxial $(La,Pr)_3Ni_2O_7$ thin films grown under compressive strain on $SrLaAlO_4$ substrates. This $T_c$ leap is enabled by pushing our gigantic-oxidative atomic-layer-by-layer epitaxy (GAE) method into an extreme non-equilibrium growth regime. It simultaneously enhances kinetics via higher temperatures and achieves full oxygenation *in situ* without post-annealing. Synchrotron X-ray diffraction and scanning transmission electron microscopy confirm that this approach yields films of large-scale crystalline purity, overcoming the inherent metastability of the strained superconducting phase. Transport measurements reveal a zero-resistance temperature ($T_c^{zero}$) reaching ~37 K, while mutual inductance measurements demonstrate a robust diamagnetic transition starting at ~23 K. These films exhibit a systematic evolution in their normal-state resistivity-temperature curve: the power-law exponent $\alpha$ evolves from Fermi-liquid-like ($\alpha$ ~2) at lower $T_c^{onset}$ to strange-metal-like ($\alpha$ ~1) in higher $T_c^{onset}$ samples, directly linking the enhanced superconductivity to non-Fermi liquid behavior. Mapping the vortex melting phase diagram by the mutual inductance technique further reveals 2D melting limit suppressed to near zero, which demonstrates significantly stronger interlayer coupling than that of cuprates. These results identify the nickelates as an ambient-pressure strange-metal high-temperature superconductors with strong interlayer coupling.

## INTRODUCTION

The century-long pursuit of higher superconducting transition temperatures ($T_c$), from conventional alloys [1] to cuprate [2-4] and iron-based [5-7] systems, has been largely propelled by the exploration of chemical compositions and doping. This successful strategy relies on the thermodynamic stability of the target compounds, which allows for robust synthesis in bulk form. The emerging family of nickelate superconductors [8,9], however, marks a departure from this paradigm: their superconducting phases are metastable, necessitating stabilization via pressurization or epitaxial strain, and exhibit an exquisite sensitivity to oxygen stoichiometry far beyond that of cuprates. The path to higher ambient-pressure $T_c$ in nickelates shifts from a mere search for chemical compositions to an act of atomic-scale engineering, where precise control over the epitaxial lattice, interfacial strain, and oxygenation becomes the decisive factor.

To date, the ambient-pressure onset transition temperature ($T_c^{onset}$) of the square-planar nickelates reached the 40 K McMillan limit [10-13], whereas the Ruddlesden-Popper (RP) bilayer nickelates [14-17] (Fig. 1a) only marginally surpasses it—still behind the records of iron-based (~55 K), cuprate (~133 K), and the RP phase itself under pressure (~96 K) (ref. [18-23]). Particularly, thin-film Meissner diamagnetism that characterizes global phase coherence has been limited under 10 K (ref. [14-16]). This performance gap partially arises from a fundamental thermodynamic dilemma in the synthesis of superconducting RP bilayer films: the hyper-oxygenated state essential for superconductivity is thermodynamically incompatible with the stability of the bilayer crystalline phase [24]. To decouple this conflict, a two-step strategy is conventionally employed, in which an insulating film is first grown for structural stability and then aggressively post-annealed to force the oxygenation required for superconductivity, at the risk of compromising structural quality (Supplementary Data Fig. S1).

This synthesis dilemma has, in turn, obscured the intrinsic physics of this new material family. Understanding the superconductivity mechanism requires probing two key aspects [4,25,26]: the pairing mechanism, which is highly correlated to the normal state from which it emerges, and the superconducting phase coherence, which can be studied via vortex dynamics. In the normal state, a hallmark of unconventional superconductivity is the "strange metal" behavior (i.e., $R \propto T$), often linked to the optimized superconductivity [27]. While this is widely observed in cuprates, infinite-layer nickelates [12], and the high-pressure RP nickelates [19-21], previous ambient-pressure superconducting RP nickelate films have been dominated by Fermi-liquid-like (i.e., $R \sim T^2$) behavior [15,16], leaving it unclear if the emergence of high-$T_c$ in nickelates is tied to this non-Fermi liquid behavior. In the superconducting state, probing phase coherence is equally crucial [28]. However, structural imperfections and uncontrolled oxygen stoichiometry in previous endeavors have resulted in either a low global phase coherence temperature or inhomogeneous phase stiffness, precluding detailed measurements of the vortex dynamics.

While pulsed laser deposition (PLD) requires a narrow thermodynamic growth window analogous to bulk synthesis [16,24], the atomic-layer-by-layer nature of oxide molecular beam epitaxy (OMBE) significantly widens thermodynamic window [29-33]. However, this potential in OMBE is constrained by the mean free path of evaporated materials, which cap the chamber oxidant pressure far too low to induce superconductivity in as-grown films [17]. Here, we unlock this potential by developing our gigantic-oxidative atomic-layer-by-layer epitaxy (GAE) method [34,35], to create an extreme non-equilibrium condition (Fig. 1b). This approach combines an ultra-strong oxidizing ambient (~1000 times that of OMBE) with an elevated growth temperature (>100°C higher than both typical PLD and OMBE) to dramatically enhance surface kinetics. GAE enables a single-step synthesis of superconducting RP bilayer nickelate films (Supplementary Data Fig. S2), achieving $T_c^{onset}$ above 60 K, a zero-resistance state at ~37 K, diamagnetic signal up to ~23 K, and substantially improved crystallinity.

## RESULTS

### Superconducting Transport Properties

Figure 1c shows a representative resistivity-temperature ($R$-$T$) curve for our optimized $(La,Pr)_3Ni_2O_7$ film, grown as-synthesized without post-annealing. The S1 film exhibits a superconducting onset transition temperature ($T_c^{onset}$) of ~63 K (Fig. 1c, upper inset). We adopt the deviation from linearity, attempting to capture the initial Cooper pair formation, thereby revealing the intrinsic pairing potential of this oxygen-sensitive metastable system. The normal-state resistivity is largely linear, following the power-law relation $R(T) = cT^{\alpha} + R_0$ with an exponent $\alpha = 1.11$, close to the $\alpha = 1$ characteristic of strange metals. This linear-like $R$-$T$ behavior for samples with $T_c^{onset}$ ~60 K is rather robust against sample details, such as the slope and residual resistivity (Supplementary Data Fig. S3). Another representative sample S2 exhibit $T_c^{onset}$ of ~56 K, and apparently larger $\alpha$. Another characteristic temperature $T_1$, defined as the intersection of the normal-state and transition-edge linear extrapolations, reaches 50 K for sample S1 and 47 K for sample S2 (not shown in Fig. 1, see Supplementary Data Fig. 4 for details). At lower temperatures, the extrapolated $R$-$T$ linear fit intersects with the zero-resistance floor ($T_2$) at 44 K for S1 and 42 K for S2, and the resistance drops below the measurement noise level ($T_c^{zero}$) at approximately 30 K for S1 and 37 K for S2 (Fig. 1c, lower inset). To rigorously benchmark our results against existing representative results in the literature, we provide a comprehensive analysis of various $T_c$ definitions ($T_c^{onset}$, $T_1$, $T_c^{50\%}$, $T_2$, $T_c^{zero}$) in Supplementary Data Fig. S4 (see discussion part for physical interpretation of different characteristic temperatures).

Figure 1d shows normalized $R$-$T$ curves for three representative films with $T_c^{onset}$ varied from 60 K, 46 K, to 38 K, the power-law exponent $\alpha$ increases from 1.18, 1.59, to 1.84. This correlation is summarized in Fig. 1e. The statistical analysis of 90 high-quality samples (all with resistance dropping below the noise level at temperatures above 15 K) reveals a trend.

This dataset includes films with different La:Pr ratios (range from 67:33 to 55:45), yet all data largely collapse onto a universal curve. This demonstrates that the superconducting properties are not sensitive to small variations in the La:Pr ratio. While the isovalent (3+) Pr and La ions have different oxygen affinities, which can affect the ease of oxidation, this ratio is only one of several GAE parameters (e.g., growth temperature, oxidant pressure, cool-down, see Methods) that are tuned. The collapse onto a single curve implies that the final oxygen stoichiometry, achieved through this multi-parameter optimization, is the dominant parameter governing the $\alpha$-$T_c^{onset}$ evolution [36,37]. The highest-$T_c^{onset}$ samples with strange-metal like behavior (approaching $\alpha = 1$) represent the optimal oxidation state; as the samples become progressively over-oxidized (i.e., $(La,Pr)_3Ni_2O_{7+\delta}$), $T_c^{onset}$ is gradually suppressed and the resistivity exponent $\alpha$ systematically increases towards the Fermi-liquid-like behavior (approaching $\alpha = 2$). Notably, previous reports of ambient-pressure films [15,16] showing $\alpha \sim 2$ with $T_c^{onset} \sim 48$ K fall consistently onto this trendline, suggesting they correspond to the over-oxidized regime. This evolution establishes a link between the enhanced superconductivity and the non-Fermi liquid, strange-metal behavior, for thin films under ambient pressure.

**Vortex Dynamics**

The improved crystallinity enhances the superconducting phase coherence, enabling detailed studies of the intrinsic vortex dynamics. We employed a two-coil mutual-inductance technique to probe the magnetic field penetration across the superconducting film (Fig. 2a inset). As shown in Fig. 2a, at zero magnetic field, the real part of the pickup coil voltage exhibits a drop, while the imaginary part simultaneously shows a corresponding peak. At lower temperatures, the real part saturates and the imaginary part returns to zero, confirming the transition into a robust, diamagnetic superconducting state. We note that for thin films with thickness ~7 nm, the effective penetration depth ($\Lambda = 2\lambda^2/d$, where $\lambda$ is the bulk London penetration depth) diverges to macroscopic scales. Given the finite lateral size of the film, magnetic flux leakage around the sample boundaries is inevitable. Consequently, complete magnetic screening is geometrically precluded. Despite this limitation, the observed ~80% signal drop represents a robust global superconducting phase coherence across the superconducting film. We define the vortex melting temperature $T_M$ as the onset of the imaginary signal (marked by arrows), which is ~23 K at zero field, significantly higher than previous reports of ~10 K (ref. [15,16], see Supplementary Data Fig. S6). As an external magnetic field ($B$) is applied, $T_M(B)$ systematically shifts to lower temperatures, with a clear transition remaining visible above 5 K at 14 T.

The $T_M(B)$ data points (colored circles) form a vortex melting boundary in the $B$-$T$ phase diagram (Fig. 2b). This melting line exhibits a distinct downward curvature. This behavior is fundamentally different from the typical temperature dependence of the upper critical field, which is either linear for out-of-plane fields or shows an upward curvature for in-plane fields in the RP bilayer nickelate systems [14-17]. This characteristic concave-down shape is the hallmark of vortex melting in layered superconductors [38]. In this model, the vortex system behaves as a three-dimensional (3D) anisotropic solid at low fields. As the field increases beyond a crossover field $B_{cr}$, the interlayer Josephson coupling becomes progressively less relevant, and the vortex lines across the entire thickness decouple into independent two-dimensional (2D) vortices within each layer. This crossover from 3D to quasi-2D behavior governs the shape of the melting line. We fit our experimental $T_M(B)$ data to the equation:

$$T_{\mathrm{M}}(B) = T_{\mathrm{M}}^{2\mathrm{D}}\left[1 + \left(\frac{1}{\ln(B/B_{cr} + c)}\right)^{1/\nu}\right], \quad (1)$$

where $T_M^{2D}$ represents the theoretical limit of vortex melting temperature of isolated 2D vortex systems, the exponent $\nu = 0.37$, and $c$ is a fitting constant. The fit, shown as the solid gray line in Fig. 2b, yields a $T_M(B)$ suppressed to near zero (~0.05 K) and a massive crossover field of $B_{cr}$ ~200 T. This reveals a strong contrast to highly anisotropic cuprates such as $Bi_2Sr_2CaCu_2O_{8+d}$ (Bi-2212, dashed line in Fig. 2b, ref. [39]) and $YBa_2Cu_3O_{7-\delta}$ (generally considered having stronger interlayer coupling than Bi-2212, ref. [40]). For Bi-2212, the weak interlayer coupling leads to a small crossover field of $B_{cr}$ ~2 T and a high $T_M^{2D}$ of ~12 K. The suppression of $T_M^{2D}$ to near zero, combined with a $B_{cr}$ two orders of magnitude larger, unequivocally demonstrates that the interlayer coupling in the RP bilayer nickelate system is much stronger than in Bi-2212. This strong coupling preserves the 3D-like nature of the vortex system to much higher fields, placing the RP nickelates in a distinct vortex physics regime compared to Bi-2212.

**Structural Characterizations**

The structural integrity of the films is corroborated by detailed structural and chemical analysis. Figure 3a shows large-scale high-angle annular dark-field (HAADF)-scanning transmission electron microscopy (STEM) images taken from three different regions of the same sample, each covering a horizontal span greater than 80 nm. The images demonstrate uniformity and sharp, well-defined interfaces over wide areas. Crucially, we observe no intergrowths of adjacent RP phases (e.g., layer number $n$=1 or $n$=3), a common challenge for this metastable system, confirming the phase purity of our films. This long-range structural order is particularly noteworthy. Given the inherent metastability of the RP bilayer nickelate phase, achieving such crystalline coherence over extended areas signifies a substantial advance in synthesis methodology.

At the atomic scale, HAADF and corresponding energy-dispersive X-ray spectroscopy (EDS) elemental maps (Fig. 3b) provide direct visualization of the interfacial structure. The Al signals from the substrate and Ni signals from the buffer layer, validating the as-designed Ni-Al-O bilayer interface structure depicted in Fig. 1a. This engineered interface is critical, as it serves as a stable template for the subsequent bilayer growth, eliminating the need for substrate surface treatments previously essentail for superconducting films [15,17].

We further investigated the film's crystallographic structure using high-resolution synchrotron X-ray diffraction (XRD) and reciprocal space mapping (RSM), as shown in Fig. 4. The out-of-plane scan (Fig. 4a) for a 3 unit cell (UC) $(La,Pr)_3Ni_2O_7$ film reveals a series of sharp (00*L*) film peaks. Pronounced Laue oscillations are visible surrounding the main Bragg peaks, which is a definitive signature of atomically sharp interfaces, and uniform film thickness. A fit to these oscillations yields a film thickness of 6.98 nm, which is consistent with the designed 3 UC $(La,Pr)_3Ni_2O_7$ structure plus the buffer layer. To determine the in-plane strain state and symmetry, we performed RSMs around the $SrLaAlO_4$ (10$\underline{11}$) diffraction peak (Fig. 4b). The film's corresponding (10$\underline{17}$) peak is observed at the exact same in-plane reciprocal space coordinate ($q_x$) as the substrate peak. This vertical alignment provides evidence that the film is coherently strained to the substrate. Furthermore, the RSMs collected at four different in-plane rotation angles ($\Phi$ = 0°, 90°, 180°, 270°) show the film peak at identical $q_x$ and $q_z$ positions, confirming the film's in-plane tetragonal symmetry inherited from the substrate. The crystalline quality is further highlighted by a direct comparison to previous work [15] (Supplementary Data Fig. 5), where both the (00*L*) diffraction peaks and the (10$\underline{17}$) RSM profile are substantially more intense.

**DISCUSSION**

The GAE method differs thermodynamically and kinetically from conventional ozone PLD or MBE. In the thermodynamic aspect, both the atomic-layer-by-layer growth mode and the enhanced oxidation power greatly expands the thermodynamic stability window, enabling a higher-temperature growth regime inaccessible to ozone PLD or MBE, where the phase would otherwise decompose. In the kinetic aspect, the elevated thermal energy (>100°C higher than previous reports), combined with the inherent high kinetic energy of the laser-ablated plasma (unlike evaporative MBE) depositing at atomic-layer level, creates an extreme kinetic environment. This allows for the rapid healing of defects and the establishment of long-range order while simultaneously maintaining the oxygen stoichiometry. Therefore, GAE creates an extreme non-equilibrium growth regime that effectively decouples the conflict between crystallinity and oxygenation.

The advancement over our initial report [15] stems from two synergistic optimizations: the engineering of a coherent interface reconstruction using a monolayer nickel-oxide buffer [26], and the implementation of single-step GAE process with elevated temperature. In contrast to conventional two-step approaches that reply on low-temperature topotatic oxidation, we hypothesize that our high-temperature oxidation (~850°C) during growth followed by rapid quenching (~100°C/min) accesses a unique thermodynamic trajectory. It likely enables the saturation of essential in-plane sites while suppressing disordered interstitial accumulation, thereby locking in an optimal oxygen configuration that supports higher-$T_c$ superconductivity.

The *R-T* curves of the RP bilayer nickelates shown in this work and previous reports [14-17] feature broader transition than typical curpate and iron-based counterparts (Supplementary Data Figure 4). This broadening likely reflects the sensitivity of phase coherence to the specific oxygen distribution in these metastable systems. In these *R*-*T* curves, the initial deviation from the normal-state resistivity ($T_c^{onset}$) may signify the onset of Cooper pairing, corroborated by recent angle-resolved photoemission spectroscopic evidence linking the superconducting gap formation to the resistive onset [41]. As the temperature decreases, the characteristic temperature $T_1$ could correspond to the emergence of locally phase-coherent superconducting puddles. The subsequent transition region from $T_1$ to $T_2$ (passing through the intermediate $T_c^{50\%}$) likely reflects a percolation process, where these islands might expand and couple via Josephson tunneling [42]. In this scenario, the zero-resistance state ($T_c^{zero}$) would be reached when the first continuous superconducting path traverses the sample, whereas the onset of the Meissner effect ($T_M$) suggests the establishment of global phase coherence throughout the film volume. This gradual establishment of macroscopic superconducting phase coherence might be intrinsically modulated by the oxygen sublattice: planar oxygen vacancies tend to disrupt the in-plane phase coherence (particularly relevant to the $d_{x2-y2}$ orbital), apical oxygen defects weaken the $d_{z2}$-mediated interlayer coupling, while interstitial oxygens lead to extra hole doping [37]. Therefore, our definition of resistive onset plausibly captures the inherent 60-K-class pairing energy scale of the ambient-pressure RP bilayer nickelates. It is likely for this reason, namely that $T_c^{onset}$ reflects the intrinsic pairing energy scale, that we observe its strong correlation with the normal-state strange-metal exponent $\alpha$. Crucially, the mutual inductance measurement highlights the macroscopic quality of our superconducting state compared to representative previous reports (see Supplementary Data Fig. S6).

In summary, by pushing the epitaxial synthesis into an extreme non-equilibrium regime, we have resolved the fundamental thermodynamic dilemma between phase stability and hyper-oxygenation. This approach enables the single-step growth of films that overcomes the inherent metastability of the superconducting phase, achieving substantially improved structural integrity over large areas and an onset superconducting transition temperature up to 63 K, and a zero-resistance temperature up to 37 K. These films reveal a link between the enhanced $T_c$ and linear-*T* strange-metal behavior. Enabled by improved diamagnetism starting at 23 K, mutual inductance measurements reveal a robust vortex state with strong interlayer coupling. Our work demonstrates that by overcoming thermodynamic constraints via GAE, ambient-pressure nickelates can achieve pairing energy scales approaching the cuprate regime. The observed evolution from Fermi-liquid to strange-metal behavior, combined with strong interlayer coupling, establishes the RP nickelates as an intriguing platform for investigating the origin of high-$T_c$ pairing.

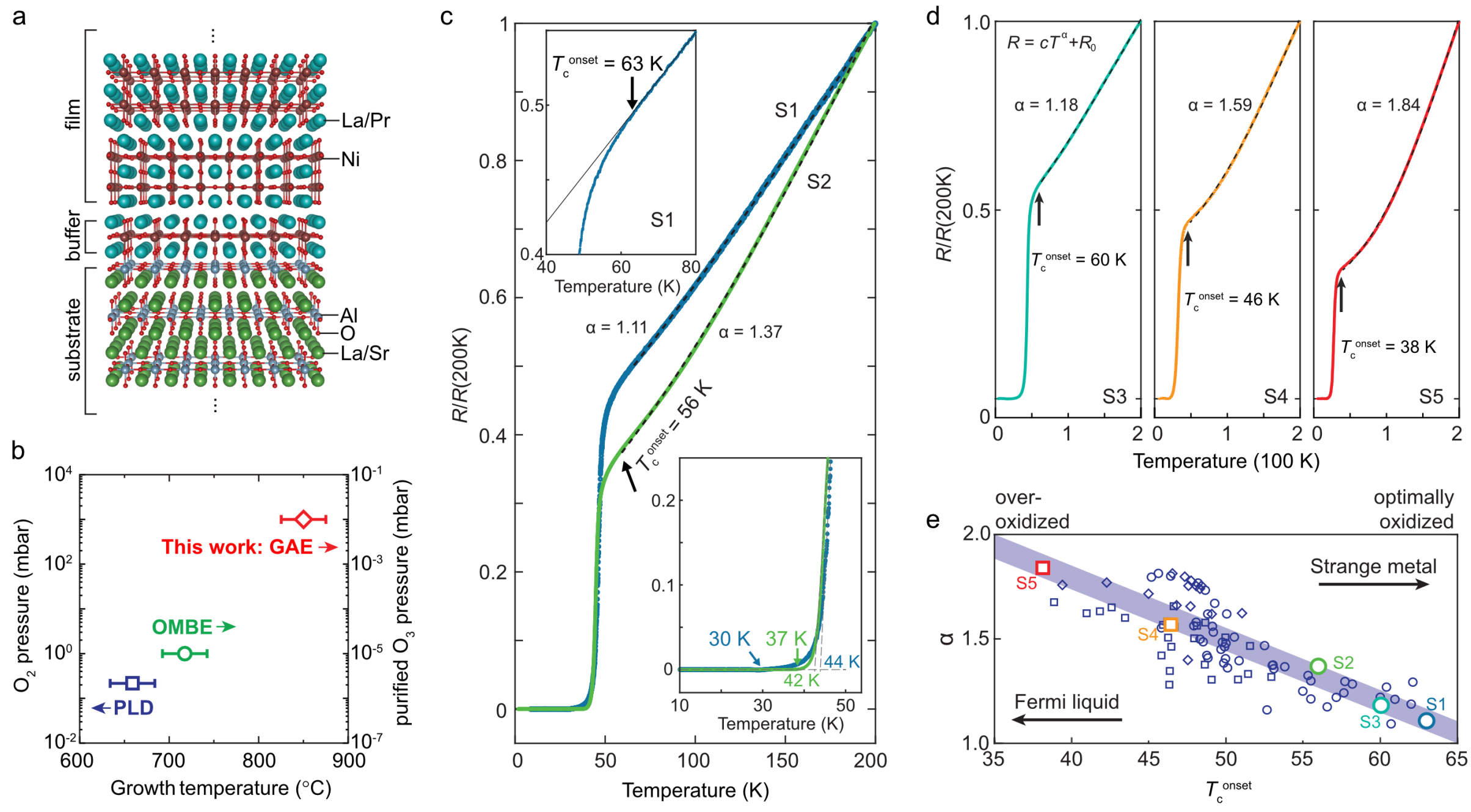

**Figure 1. Synthesis and transport properties of ambient-pressure bilayer nickelate superconductors with $T_c$ over 60 K. (a)**, Structural schematic of the $(La,Pr)_3Ni_2O_7$ bilayer film, grown on a (001)-oriented $SrLaAlO_4$ substrate and buffered by a monolayer nickelate, forming a Ni-Al-O bilayer structure at the interface. **(b)**, Growth parameter space (growth temperature versus oxidant pressure) for our gigantic-oxidative atomic-layer-by-layer epitaxy (GAE) method in this work, contrasted with conventional pulsed laser deposition (PLD) and oxide molecular beam epitaxy (OMBE). GAE accesses a unique high-temperature and high-oxidation regime, enabling extreme non-equilibrium synthesis. **(c)**, Representative resistivity-temperature (*R*-*T*) curves for two $(La,Pr)_3Ni_2O_7$ films (sample S1 and S2). The normal state parts (dashed lines) are fit to $R = cT^{\alpha} + R_0$ with exponents $\alpha = 1.11$ and $\alpha = 1.37$. Upper inset: Zoom-in of the S1 transition, where $T_c^{onset} = 63$ K is defined as the deviation from a linear-in-*T* fit (gray line). $T_c^{onset} = 56$ K for S2. Lower inset: Zoom-in near zero resistivity, showing the extrapolated linear fits intersecting the zero-resistance axis at 44 K for S1 and 42 K for S2. The resistance drops to the measurement noise level at approximately 30 K for S1 and 37 K for S2. **(d)**, Normalized *R*-*T* curves (*R*/*R*(200 K)) for three films with corresponding to $T_c^{onset}$ values of 60 K, 46 K, and 38 K. Dashed lines are power-law fits, yielding exponents $\alpha$ = 1.18 , 1.59, and 1.84, respectively. **(e)**, The normal state resistivity exponent $\alpha$ plotted as a function of $T_c^{onset}$ for 90 samples with zero resistance above 15 K. The symbols denote different $(La,Pr)_3Ni_2O_7$ films with varying La:Pr ratios: 67:33 (squares), 65:35 (circles), and 55:45 (diamonds). The highlighted circles and squares (S1–S4) correspond to the samples with *R*-*T* plotted in respective colors shown in (c) and (d).

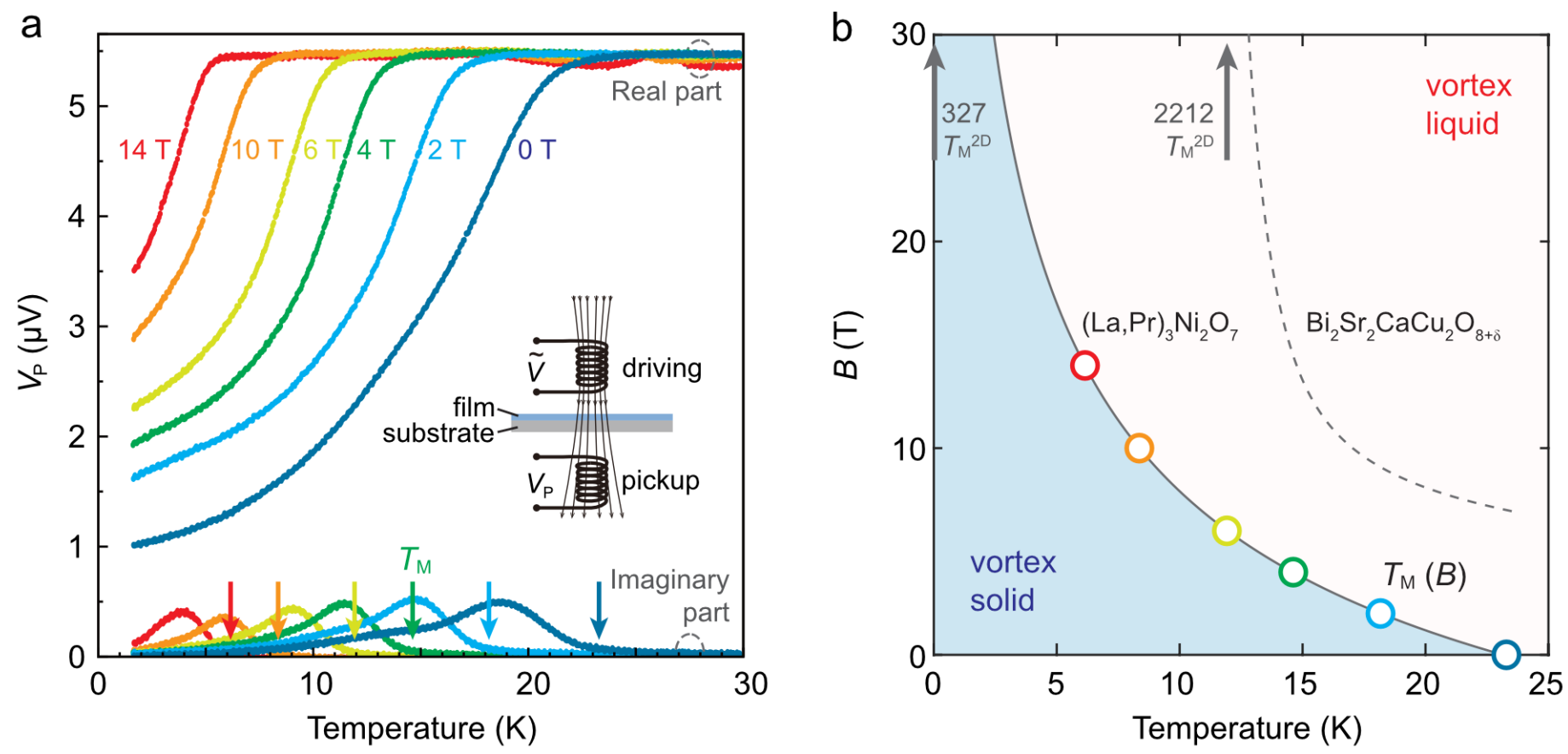

**Figure 2. Vortex melting phase diagram of the bilayer nickelate superconductor. (a)**, Temperature dependence of the real (upper curves) and imaginary (lower curves) components of the pickup coil voltage ($V_P$), measured using a two-coil mutual-inductance technique (inset schematic) under various magnetic fields. The real part indicates the diamagnetic shielding. The onset of the imaginary signal, marked by the arrows, defines the vortex melting temperature ($T_M$). **(b)**, The magnetic field-temperature phase diagram, plotting the $T_M (B)$ points extracted from the onsets in (a) (colored circles). The solid gray line is a theoretical fit. This line separates the vortex solid phase from the vortex liquid phase. For comparison, the dashed gray line shows the melting line for the highly anisotropic cuprate $Bi_2Sr_2CaCu_2O_{8+\delta}$ (2212, data adapted from ref. [20]). The upward-pointing arrows indicate the respective 2D melting temperatures ($T_M^{2D}$, yielded from theoretical fits), the high-field limit where interlayer coupling vanishes. In contrast to the finite $T_M^{2D}$ (~12 K) of 2212, the 2D melting limit for $(La,Pr)_3Ni_2O_7$ (327) is suppressed to near zero. This demonstrates the much stronger interlayer coupling in the nickelate system, highlighting its distinct vortex dynamics.

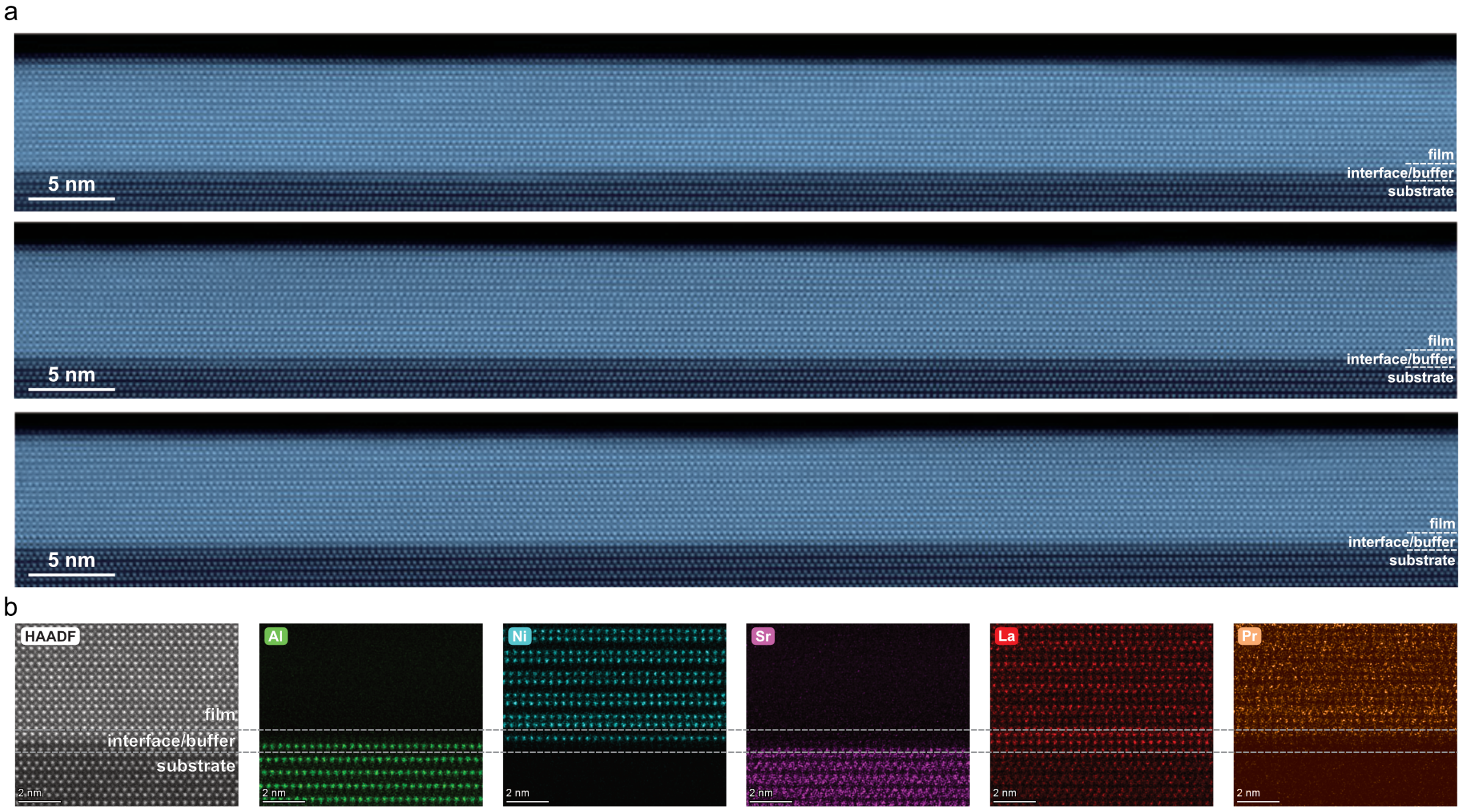

**Figure 3. Large-scale structural and chemical characterization of a superconducting bilayer nickelate thin film with $T_c^{onset}$ over 60 K. (a),** Large-scale high-angle annular dark-field (HAADF)-scanning transmission electron microscopy (STEM) images taken from three different regions of the same sample, demonstrating uniform crystalline structure and interfacial sharpness over wide areas. **(b)**, Atomic-resolution HAADF-STEM image and corresponding energy-dispersive X-ray spectroscopy (EDS) elemental maps (Al, Ni, Sr, La, Pr) of the interface, confirming the formation of Ni-Al-O bilayer interface structure depicted in Fig. 1a.

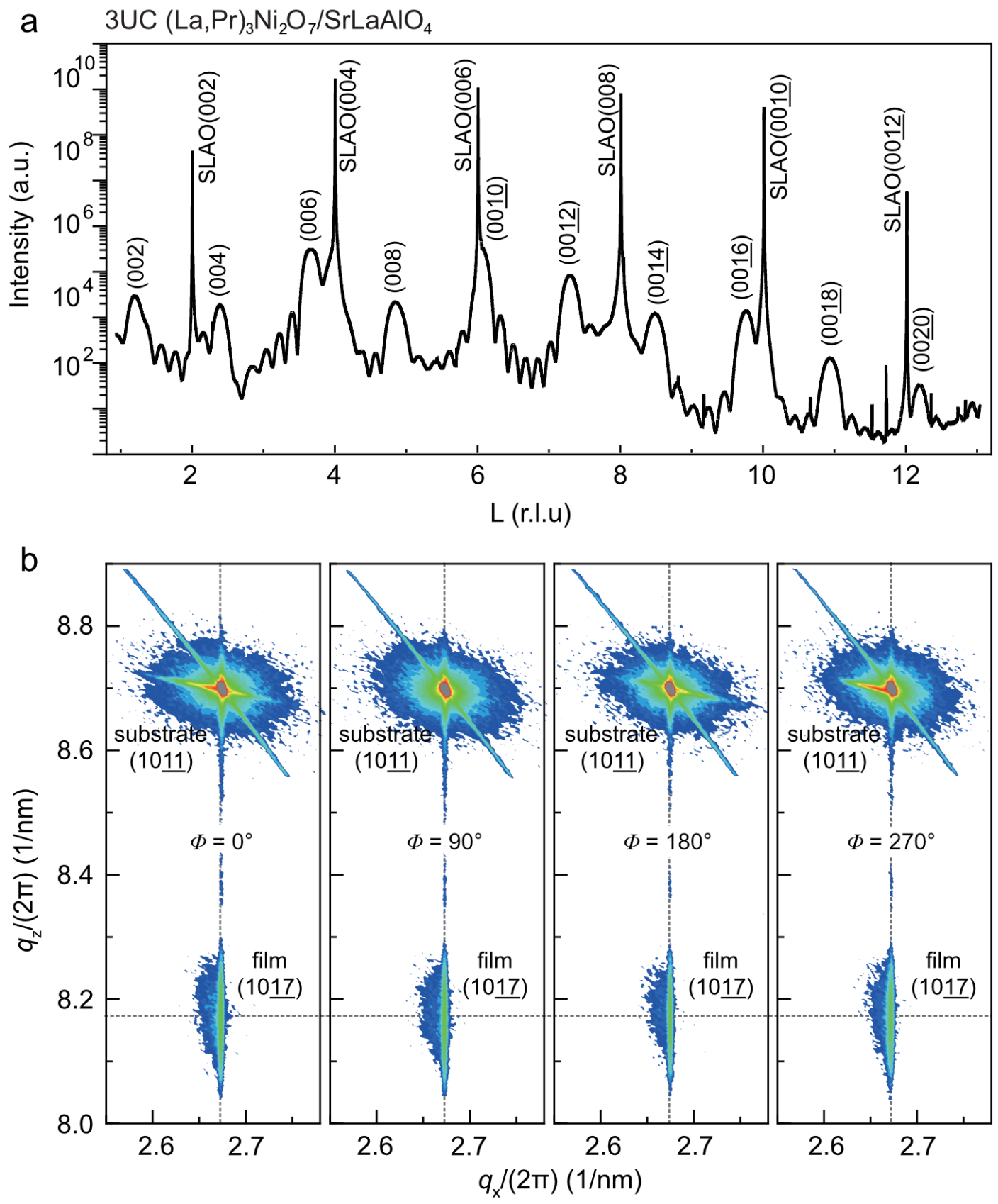

**Figure 4. X-ray diffraction (XRD) and reciprocal space mapping (RSM) of a superconducting bilayer nickelate thin film with $T_c^{onset}$ over 60 K. (a),** High-resolution synchrotron XRD out-of-plane scan for a 3 unit cell (UC) $(La,Pr)_3Ni_2O_7$ film. The spectrum displays film diffraction peaks from (002) to (00$\underline{20}$) alongside $SrLaAlO_4$ (SLAO) substrate peaks. Several sharp spikes are observed, which originate from minor structural imperfections within the bulk of the single-crystal substrate in response to the intense synchrotron X-rays, and are irrelevant to the film. **(b)**, RSMs at four different in-plane rotation angles ($\Phi$ = 0°, 90°, 180°, 270°). The corresponding film (10$\underline{17}$) peaks are observed at the exact same in-plane momentum ($q_x$) as the substrate (10$\underline{11}$) peaks, confirming that the film is coherently strained. The identical positions of the film peak across all four $\Phi$ angles demonstrate the film's in-plane tetragonal symmetry.

## METHODS

**Growth of $(La,Pr)_3Ni_2O_7$ films.** Using the gigantic-oxidative atomically layer-by-layer epitaxy (GAE) method [6,19], by alternately ablating the $(La,Pr)O_x$ and $NiO_x$ targets with pulsed laser, the $(La,Pr)_3Ni_2O_7$ films with a stacking block structure of $(La,Pr)O$-$NiO_2$-$(La,Pr)O$-$NiO_2$-$(La,Pr)O$ were grown. On as-received polished $SrLaAlO_4$ substrates (MTI-Kejing), a $(La,Pr)O$-$NiO_2$-$(La,Pr)O$ buffer is firstly grown to mitigate interfacial structural discontinuity [25]. The buffer layer is insulating. During the growth process, by precisely controlling the number and energy of the pulsed lasers, combined with the oscillation curve of the reflection high-energy electron diffraction (RHEED), the complete layer coverage of the atomic layer can be determined, achieving a high stoichiometric accuracy. In a typical growth process, the number of pulses used to ablate the $(La,Pr)O_x$ or $NiO_x$ target is between 100 and 150 pulses, and the laser fluence is set at 1.4–1.8 J/cm$^2$. The chamber pressure during deposition is 10 Pa, consisting of 1–2 Pa of purified ozone and 8–9 Pa of oxygen. A water-cooled ozone nozzle (inner diameter of 10 mm) is specially designed to aim directly at the substrate and is closely positioned (~4 cm) to establish a concentrated zone near the substrate. The substrate temperature for growth is 800–850°C, measured from the back side of the flag-type Inconel sample holder. After deposition, the samples were cooled down at a rate of 100°C/min until it reaches temperatures lower than 200°C in the same oxidative environment, before transferring to the ultrahigh vacuum load lock chamber. The final oxygen stoichiometry in the grown films is tuned by making subtle adjustments to the La:Pr ratio, growth temperature, oxidant environment, and subsequent cool-down procedures. All the samples have not undergone post-annealing treatment.

**Electrical transport measurements.** Electric-transport measurements were conducted in a closed-cycle, helium-free cryostat with a base temperature of ~1.8 K. Hall-bar contacts (Pt) were defined by magnetron sputtering through a pre-patterned hard shadow mask aligned on 5 × 5 mm$^2$ samples. During electrode deposition within the vacuum sputtering chamber, samples are cooled using liquid nitrogen to prevent oxygen loss. Standard four-terminal alternating current (AC) lock-in techniques were employed (5 μA at ~10 Hz).

**Mutual inductance measurements.** The superconducting thin-film samples were clamped directly between collinear drive and pickup coils. Both coils are wound from 30 μm-diameter wire to identical specifications: 1.5 mm outer diameter, 0.5 mm inner diameter, ~2.5 mm height and 800 turns, giving a self-inductance of ~1 mH. A 20 kHz, 10 μA alternating current supplied to the drive coil induces a voltage in the pickup coil that is detected with a lock-in amplifier.

**Scanning transmission electron microcopy (STEM).** STEM HAADF imaging of $(La,Pr)_3Ni_2O_7$ film was performed on an FEI Titan Themis G2 microscope equipped with a double spherical-aberration corrector (DCOR) and a high-brightness field-emission gun (X-FEG) with a monochromator to enhance image resolution and contrast. The inner and outer collection angles ($\beta$1 and $\beta$2) for the STEM HAADF images were set to 90 and 200 mrad. The semi-convergence angle of 25 mrad was used for both HAADF imaging and EDS chemical analyses, with the beam current adjusted to about 40 pA. EDS data were acquired in STEM mode using the Super X System. The microscope was operated at 200 kV with a beam current of 200 pA; the scan parameters were a ∼1-Å probe, ∼9 $Å^2$ scan pixel size (applied 16 ×16 sub-pixel scan), and 300 µs dwell time per pixel to minimize radiation damage. The cross-section STEM specimens were prepared using an FEI Helios 600i dual-beam FIB/scanning electron microscope (SEM). Before lift-out and final thinning, the sample surface was protected by electron beam-deposited platinum and ion beam-deposited carbon to prevent ion beam damage. All procedures were carried out at room temperature.

**X-ray diffraction (XRD).** The lattice parameters of the thin films were characterized using an automated multipurpose X-ray diffractometer (SmartLab, Rigaku Corporation), including θ-2θ scans and reciprocal space mappings (RSMs). These data were collected using a Hypix-3000 2D detector. In-plane sample alignment was performed before rotating the $\Phi$ axis to measure the RSMs around (1 0 11), (0 1 11), (-1 0 11) and (0 -1 11) $SrLaAlO_4$ Bragg reflections. Synchrotron radiation X-ray diffraction was performed at the BL02U2 beamline station of the Shanghai Synchrotron Radiation Facility (SSRF) with λ = 0.118 nm and a spot size of 80 µm × 160 µm. To avoid detector overexposure, a segmented data acquisition mode was used, and an appropriate amount of attenuation filter was inserted close to the substrate diffraction peak positions. The measured XRD intensity was corrected for attenuation by applying a corresponding factor.

**SUPPLEMENTARY DATA**

Supplementary data are available at *NSR* online.

**ACKNOWLEDGEMENTS**

We thank the staff from BL02U2 beamline of SSRF for assistance of XRD data collection. We acknowledge the support from the Station of Quantum Materials.

**FUNDING**

This work was supported by the Guangdong Major Project of Basic Research (2025B0303000004), the Natural Science Foundation of China (92565303, 12504161, 12504166, 12374455, 92265112, and 52388201), the National Key R&D Program of China (2022YFA1403101 and 2024YFA1408101), the Guangdong Provincial Quantum Science Strategic Initiative (GDZX2501001, GDZX2401004, and GDZX2201001), the Shenzhen Science and Technology Program (KQTD20240729102026004), and the Shenzhen Municipal Funding Co-Construction Program Project (SZZX2301004 and SZZX2401001).

**AUTHOR CONTRIBUTION**

Q.K.X. and Z.C. supervised the project. Z.C. initiated the study and coordinated the research efforts. G.Z. performed thin film growth with Y.C., F.P., and W.W.'s assistance. A subset of the samples is grown by W.L. and Z.N. H.W. performed low-temperature measurements and analysis. H.H. performed structural measurements and analysis. Z.C. wrote the manuscript with G.Z., H.W., and H.H.'s input. G.Z., H.W., and H.H. contributed equally to this work.

***Conflict of interest statement.*** None declared.